\documentclass[a4paper]{article}

\usepackage[utf8]{inputenc}

\usepackage{multicol}
\usepackage{amsthm}
\usepackage{bm}
\usepackage{physics}

\usepackage[dvipdfmx]{graphicx}

\usepackage{bmpsize}
\usepackage{amssymb}
\usepackage{amsmath}
\usepackage{color}
\usepackage{amsthm}
\usepackage{comment}

\usepackage{dcolumn}

\usepackage{mathptmx}

\newcommand{\BA}{\begin{eqnarray}}
\newcommand{\EA}{\end{eqnarray}}

\definecolor{dgreen}{rgb}{0.0, 0.5, 0.0}

\begin{document}

\fontsize{14pt}{16.5pt}\selectfont

\begin{center}
\bf{
Construction of topological representation of geometric patterns using Cantor self-similar set
}
\end{center}
\fontsize{12pt}{11pt}\selectfont
\begin{center}
Shousuke Ohmori$^{1,2 *)}$, 
Yoshihiro Yamazaki$^{3,4}$,
Tomoyuki Yamamoto$^{3,4}$, and \\
Akihiko Kitada$^4$\\ 
\end{center}

\bigskip

\fontsize{11pt}{14pt}\selectfont\noindent

\noindent
$^1$\it{National Institute of Technology, Gunma College, Maebashi-shi, Gunma 371-8530, Japan}\\
$^2$\it{Waseda Research Institute for Science and Engineering, Waseda University, Shinjuku, Tokyo 169-8555, Japan}\\
$^3$\it{Faculty of Science and Engineering, Waseda University, Shinjuku-ku, Tokyo 169-8555, Japan}\\
$^4$\it{Institute of Condensed-Matter Science, Comprehensive Resaerch Organization, Waseda University, Shinjuku-ku, Tokyo 169-8555, Japan}\\

\noindent
*corresponding author: 42261timemachine@ruri.waseda.jp\\
~~\\

\rm
\fontsize{11pt}{14pt}\selectfont\noindent

\baselineskip 18pt

\noindent
{\bf Abstract}\\
%
%
Universal representation of geometric patterns of disordered matters is investigated with the aid of general topology.
By utilizing the result obtained in the previous study (S. Ohmori, et.al., Phys. Scr. 94, 105213 (2019)) that any patterns can be represented by a specific topological space, a construction of topological representation of patterns using Cantor set is shown.
The obtained topological representations are then demonstrated by the contractions that characterize the self-similarity of Cantor set.
For some practical geometric patterns, e.g., network, dendritic, and clusterized patterns, their  topological representations are focused on.

\bigskip

\bigskip




%
%

\section{Introduction}
\label{sec:1}

A variety of geometric patterns found in solid and liquid states have been hugely studied from the viewpoint of the physics of disordered systems\cite{Cusack,Strobl}. 
The network (graphic) structures of polymers\cite{Tezuka}, the clusterized structure of molecular liquid\cite{Katayama,Morishita}, and the dendrite found in the process of solidification\cite{Kurz} are practical and familiar examples.
In the study of patterns of disordered matter, there is a historically significant issue of finding a mathematical description that can universally and uniformly express these geometric patterns, similar to the group theory in crystallography.
To attempt this problem, several mathematical methods based on topological property have been developed.  
For instance, the persistent homology method that is based on the algebraic topology have been applied to the study of classification of geometric structures formed by amorphous materials\cite{Hirata,Hiraoka}. 

As the other approach, 
we have studied materials structure from the viewpoint of the most fundamental topology, namely, general topology\cite{Kitada2005,Kitada2007,Kitada2016,Ohmori2019}. 
%
%
In our studies, the geometric patterns are discussed in the context of Continuum Theory,
which is one of the field of general topology\cite{Nadler}.
A topological space is called a continuum if it is a connected compact Hausdorff-space, and each geometric pattern is specified by using  continua.
For instance, a dendritic pattern is described as the topological dendrite that is a locally-connected continuum without simple closed curve.
(For details of this description, see Sec. 2.)
The continua corresponding to the geometric patterns can be expressed indirectly as the formation of a set of equivalence classes for a specific topological space with the 0-dim, perfect, compact properties.
Here, the collection of subsets of a topological space relative to equivalence classes is called a decomposition space
\footnote{
Let $(A,\tau)$ be a topological space with a topology $\tau$. A partition $\mathcal{D}$ of $A$ is a set $\{D \}$ of nonempty subsets of $A$ such that $D\cap D' = \emptyset$ for $D \not = D', D,D'\in \mathcal{D}$ and $\bigcup \mathcal{D}(=\cup_{D\in \mathcal{D}})=A$. A decomposition space $(\mathcal{D},\tau(\mathcal{D}))$ of $(A,\tau)$ is a topological space whose topology $\tau(\mathcal{D})$ on a partition $\mathcal{D}$ of $A$ is defined by $\tau(\mathcal{D}) =\{\mathcal{U} \subset \mathcal{D}; \bigcup \mathcal{U} \in \tau \}$. The space $(\mathcal{D},\tau(\mathcal{D}))$ is a kind of quotient space of $(A,\tau)$. See, for the detailed discussions, \cite{Nadler}.
}. 
Then, each pattern is associated with a decomposition space, homeomorphically, 
and hence, the patterns can be discussed uniformly through the corresponding decomposition spaces. 
%
%
%
%
%

Among our methodology, we recently proposed a mathematical model called Cantor cube model\cite{Ohmori2019}.
This model provides the practical form of the decomposition spaces of a Cantor cube $X=\{0,1\}^{\Lambda}$
\footnote{
A Cantor cube is a topological space $(X,\tau)=(\{0,1\}^{\Lambda},\tau_0^{\Lambda}),$ Card $\Lambda \geq \aleph_0$ where  $(\{0,1\}^\Lambda ,\tau_0^\Lambda )$ is the $\Lambda -$product space of $(\{0,1\},\tau_0)$ for a set $\Lambda$ and $\tau_0$ is a discrete topology for $\{0,1\}$. See for details \cite{Ohmori2019}.
}
corresponding to some geometric patterns.
For instance, 
focus on a finite graph $Y_g$ composed of arcs $E_1,\dots,E_r~(r<\infty)$.
(For the definition and the sketch of a finite graph, see Sec. 2.)
%
%
Then, the corresponding decomposition space $\mathcal{D}_g$ of $X$ can be represented as follows : 
for a node $x$ connecting with arcs $E_{t_1},\dots , E_{t_q}$ and for a point $y$ in an arc $E_i$ of $Y_g$, 
\begin{eqnarray}
		x \doteq \cup _{j=1}^q (X^{t_j} \cap S^{t_j}_x),~~ y \doteq X^{i} \cap S^{i}_y.
		\label{eqn:1-1}
\end{eqnarray}
%
%
Here, the right hand sides $\cup _{j=1}^q (X^{t_j} \cap S^{t_j}_x)$ and $X^{i} \cap S^{i}_y$ are the points of $\mathcal{D}_g$ and the sign ``$\doteq$'' shows identification of the point in $Y_g$ with the corresponding point of $\mathcal{D}_g$ homeomorphically.
%
%
%
Intuitively, $\cup _{j=1}^q X^{t_j}$ in (\ref{eqn:1-1}) represents that the point $x$ possesses just the bonds $E_{t_1},\dots , E_{t_q}$ emanated from $x$, and $S^{t_j}_{x}$ determines the position of $x$ in $E_{t_j}$ 
\footnote{
According to accurate definitions of the signs $X$ and $S$ in (\ref{eqn:1-1}), see Sec. 3 in \cite{Ohmori2019}.
}.
By the topological representation (\ref{eqn:1-1}), the network pattern $Y_g$ can be completely copied to $\mathcal{D}_g$.
%
%
In the previous study\cite{Ohmori2019}, such topological representations using  Cantor cube are found for the other geometric patterns.
However, it seems that these representations 
are too abstract to study geometric pattern even more.
In particular, the relation (\ref{eqn:1-1}) does not tell us the information on which property of Cantor cube is essential to represent the geometric pattern. 

In the present article, we consider Cantor set instead of Cantor cube.
Cantor set is a topological space homeomorphic to Cantor Middle-Third Set (abbreviated as CMTS hereafter)\cite{Vallin}.
This space has been hugely studied from viewpoints of chaos and fractal science\cite{Devaney,Feder}, because it is well known as the most typical self-similar fractal set;
its self-similarity can be characterized by equipping the contractions.  
Also, it is known that Cantor set is the practical form of Cantor cube 
\footnote{Cantor cube $(\{0,1\}^{\Lambda},\tau_0^{\Lambda})$ becomes a Cantor set when Card $\Lambda = \aleph_0$, e.g., $\Lambda =\mathbb{N}$, where $\mathbb{N}$ is the set of natural numbers.}.
%
%
%
%
The purpose of this article is to find the essential structure of topological representations by handling Cantor set.
Then, it is found that each representation is obtained as a composition of the contractions characterizing the self-similarity of Cantor set.
%
%
%

The current article is organized as follows.
In the next section, we review the classification of some geometric patterns based on continua used in our study.
%
%
In Sec. 3, we show the basic topological properties for the self-similarity of a Cantor set and we sketch the procedure of representing any pattern by a decomposition space of the Cantor set.
The obtained result in previous section is applied to some continua with graphic, dendritic, or clusterized structures in Sec. 4.
A conclusion is given in Sec. 5.


\section{Classification of geometric patterns}
\label{sec:2}

In this section, classification of geometric patterns based on the concept of continua
is reviewed.
The patterns are basically classified into the four types of topological spaces;
topological graph, topological tree, topological dendrite, and the direct sum of them.
Note that graph, tree, dendrite are continua, whereas the direct sum is not a continuum but a disconnected space.

%
\subsection{Topological Graph (Network Pattern)}
\label{sec:2-1}

\begin{figure}[h!]
    \begin{center}
	\includegraphics[width=4cm]{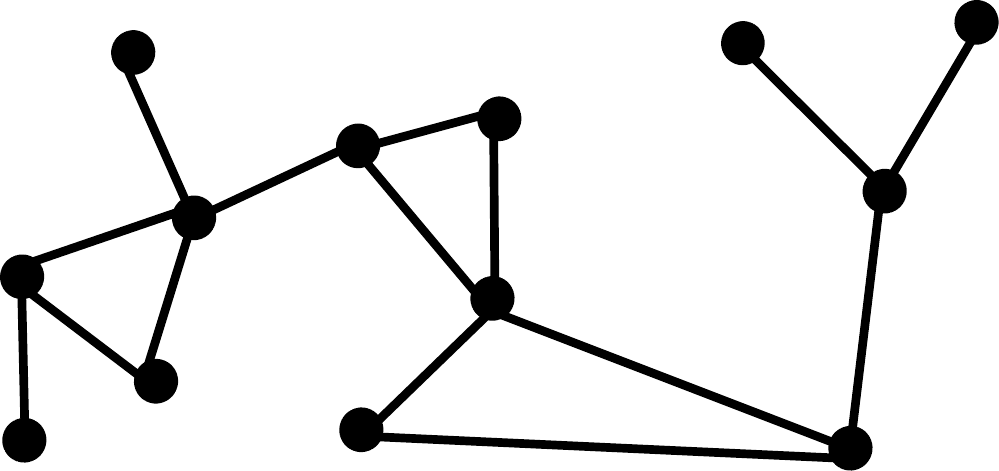}
    \hspace{7mm}
    \includegraphics[width=3cm]{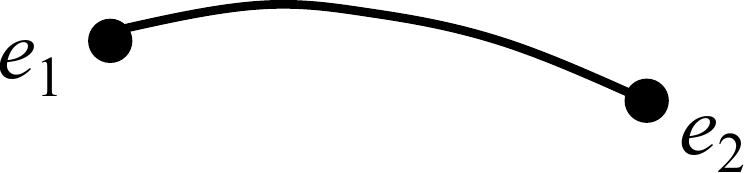}
    \hspace{7mm}
    \includegraphics[width=3cm]{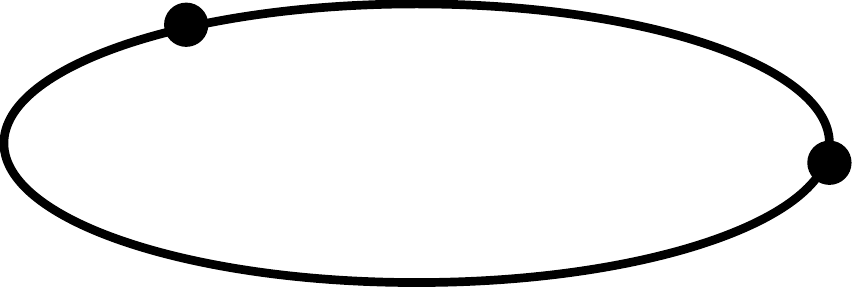}\\
    ~~~(a)
    \hspace{35mm}
    (b)
    \hspace{35mm}
    (c)
    \caption{\label{Fig.graphs} Illustrations of (a) a topological graph (network pattern),
    (b) an arc with the end points $e_1$ and $e_2$, 
    (c) a simple closed curve.
   }
   \end{center}
\end{figure}

We first focus on the geometric pattern of network (graphic) configuration.
Figure \ref{Fig.graphs} (a) shows an example of such pattern.
This pattern is widely found in many disordered matters, e.g., polymers and amorphous. 
In Continuum Theory, this pattern can be completely associated with a topological space called a topological  graph.
A topological space is called a topological graph (a finite graph, or a graph) provided that it is a continuum that can be described as the union of finitely many arcs, any two of which are either disjoint or intersect in terms of their end points
\footnote{Any graph can be also characterized by focusing on the vicinity of each point.
Set a point $x$ in a topological space $(X,\tau)$ and a cardinal number $\beta$.
We say $x$ is of less than or equal to $\beta$ in $X$, written 
\begin{equation}
    ord (x,X)\leq \beta,
    \label{eqn:2-g1}
\end{equation}
if for any open set $U\in \tau$ containing $x$, 
there exists $V\in \tau$ such that $x\in V \subset U$ 
and $Card \partial V \leq \beta$.
($\partial V$ shows the boundary of $V$.)
Then, we can prove the statement that a continuum $X$ is a graph if and only if for any $x\in X,~ord (x,X) \leq n $ for some $n\in \mathbb N$ and $ord (x,X) \leq 2$ for all but finitely many $x\in X$\cite{Nadler}.
In particular, the necessary and sufficient condition that a continuum $X$ becomes an arc or a simple closed curve is $ord (x,X) \leq 2$ for any $x\in X$.
These statements show that a topological graph can be discussed based on the abstract topological concept such as (\ref{eqn:2-g1}).
In Continuum Theory, several general topological methods are used to investigate what shapes the topological graph takes.}.
Here, an arc is a topological space homeomorphic to the closed interval $[0,1]$ and its end points are the points mapped homeomorphically into $0$ and $1$.
%
%
%
Figure \ref{Fig.graphs} (b) shows an arc in which $e_1$ and $e_2$ indicate its end points.
In practice, each arc in a graph characterizes a bond in a network pattern.
Note that any arc is itself a topological graph.
The other basic example of a graph is a simple closed curve.
This is the space  homeomorphic to $S^1:=\{x=(x_1,x_2) \in \mathbb{R}^2; |x|=\sqrt{x_1^2+x_2^2}=1\}$ 
(Fig. \ref{Fig.graphs} (c)).

\subsection{Topological Tree (Branching Pattern)}
\label{sec:2-2}

\begin{figure}[h!]
    \begin{center}
	\includegraphics[width=3cm]{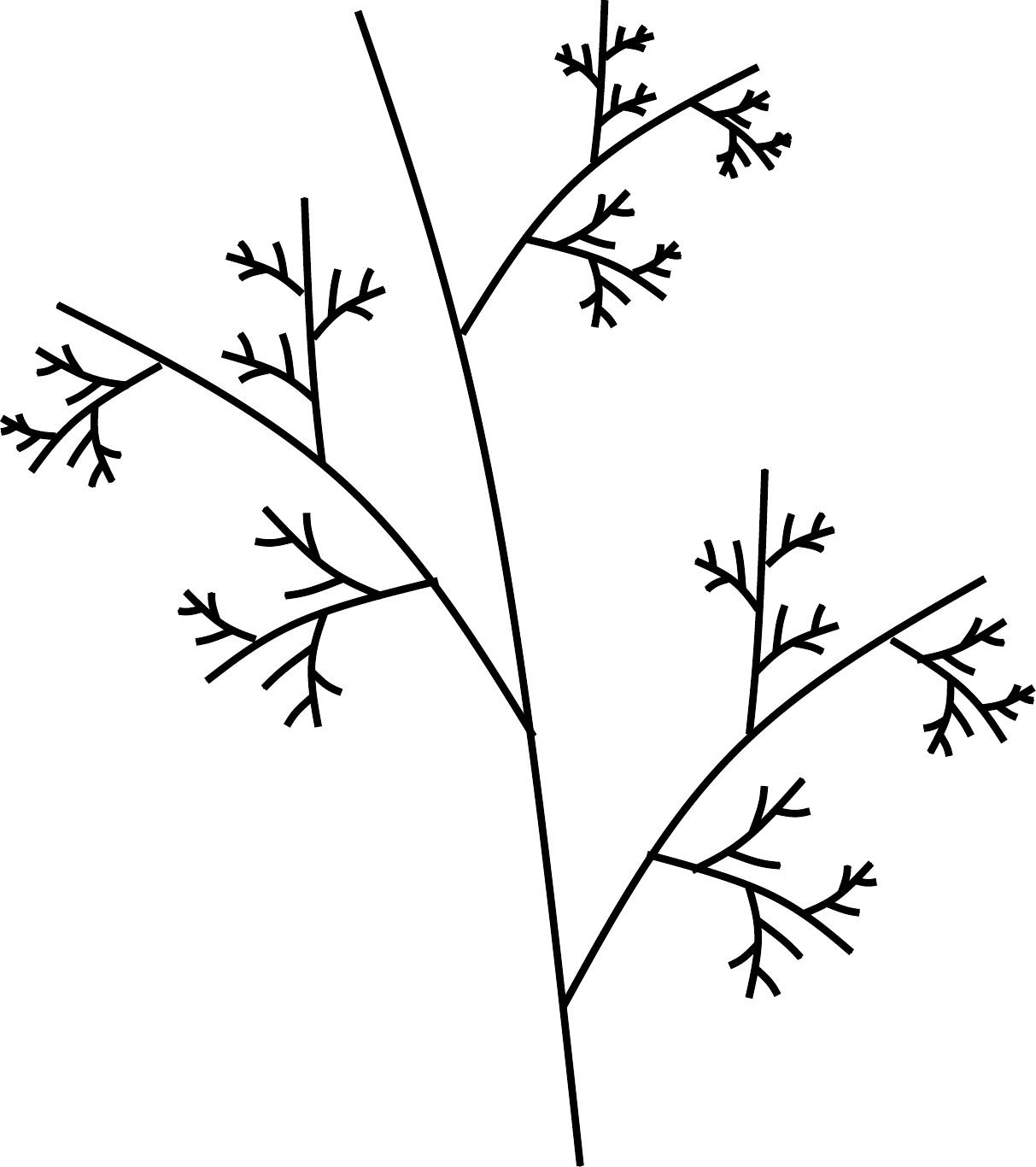}
    \hspace{13mm}
    \includegraphics[width=6cm]{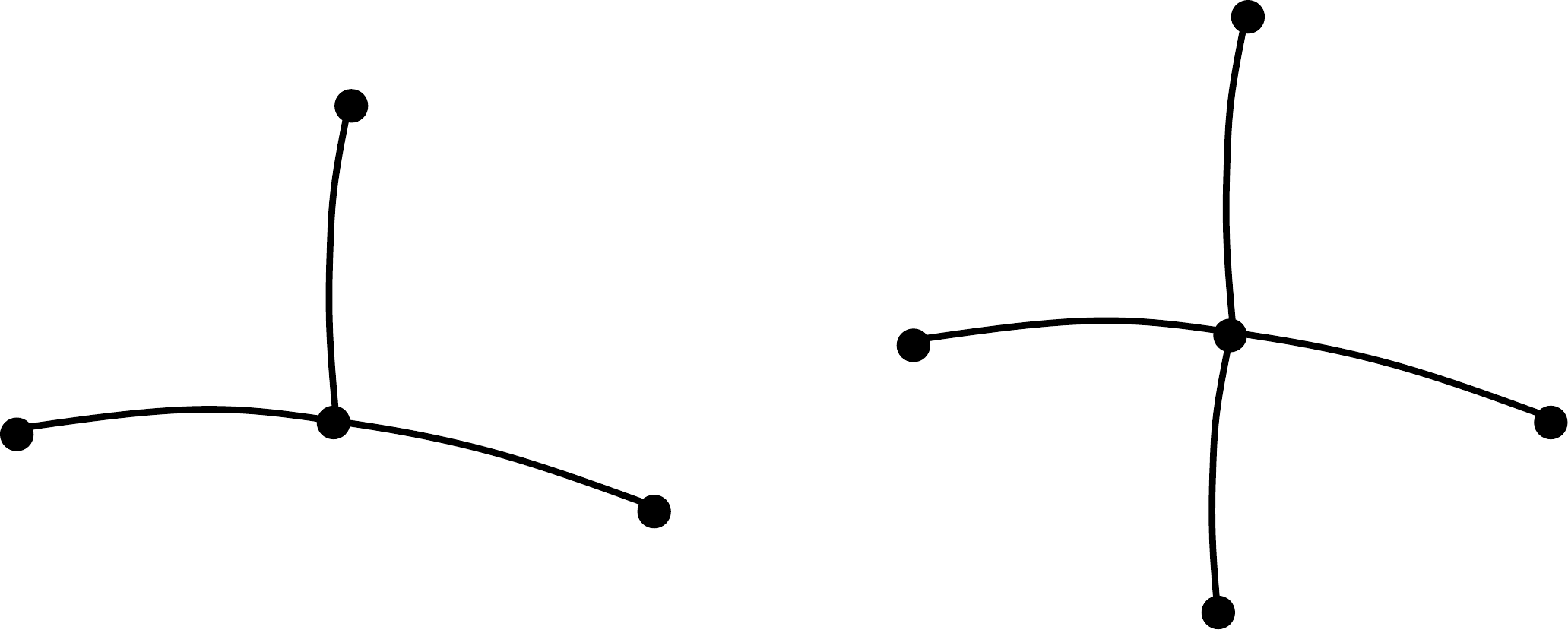}\\
    (a)
    \hspace{35mm}
    (b)
    \hspace{35mm}
    (c)
    \caption{\label{Fig.trees} Illustrations of (a) a topological tree, 
    (b) the simple $3$-od, (c) the simple $4$-od.
   }
   \end{center}
\end{figure}

The special pattern of the graph is a topological tree.
A topological tree (a tree, or an acyclic graph) is a graph that contains no simple closed curve.
Figure \ref{Fig.trees} (a) shows an example of a topological tree.
This pattern corresponds to the branching pattern such as the dendritic pattern.
The simplest example of a tree is an arc.
Note that a tree that is not an arc can be composed of the family of simple $n$-od spaces, 
where a space is a simple $n$-od ($n=3,4,\dots$) if it is a tree that has a unique point connecting with $n$ number of arcs and the other points do not connect with $l~(>2)$ number of arcs.  
Figure \ref{Fig.trees} (b) and (c) show the simple $3$-od and $4$-od spaces, respectively.
The simple $n$-od space is known as one of the key space to investigate a topological tree in Continuum Theory.  

\subsection{Topological Dendrite (Branching or Self-Similar Pattern)}
\label{sec:2-3}

\begin{figure}[h!]
    \begin{center}
	\includegraphics[width=4cm]{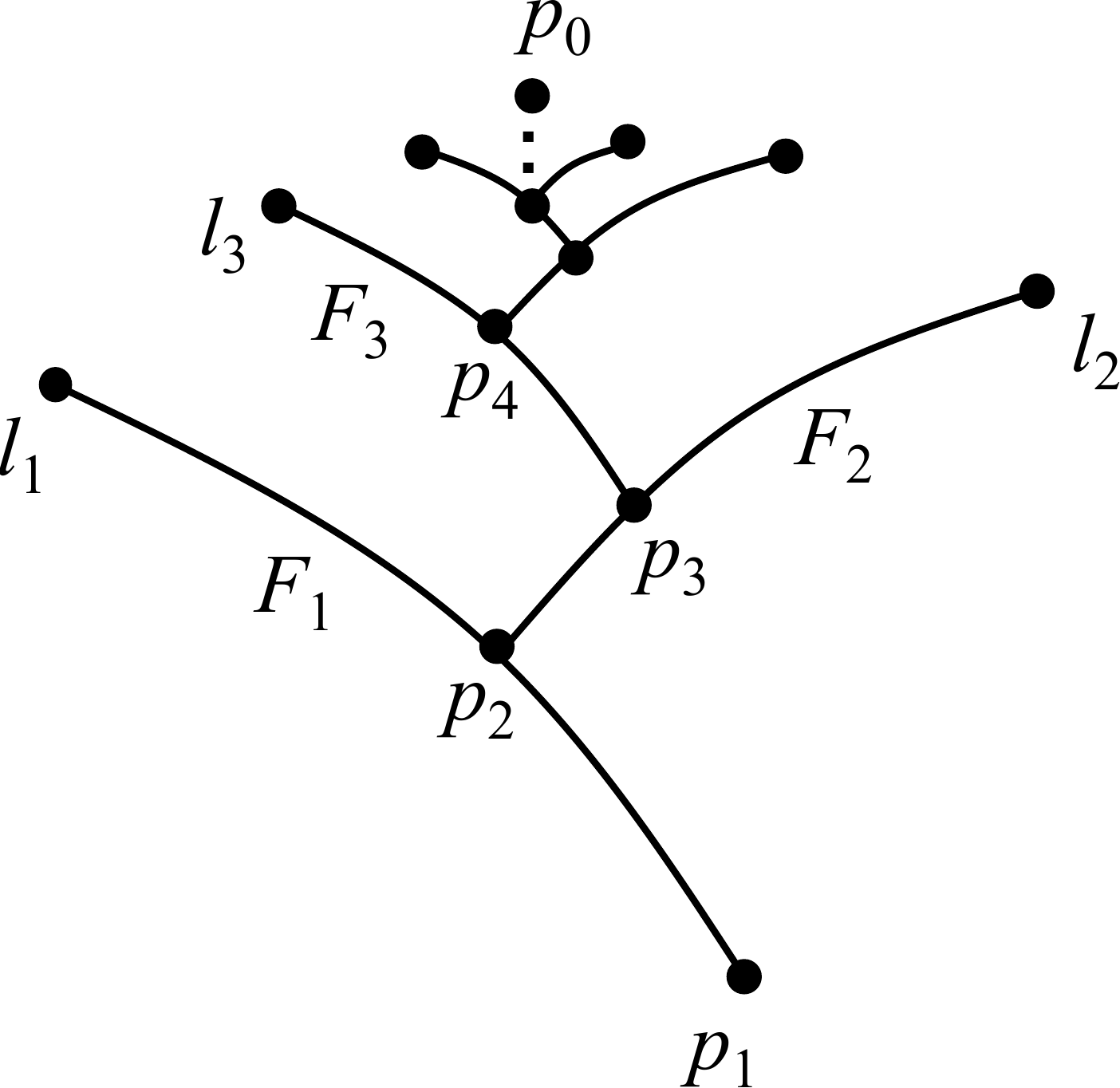}
    \hspace{13mm}
    \caption{\label{Fig.SSdend} Schematic explanation of a topological dendrite.
    $F_i$ shows an arc with the end points $p_i$ and $l_i$ ($i=1,2,\dots$).
   }
   \end{center}
\end{figure}

As a similar pattern to a tree, a topological dendrite is well-known.
This space is defined as a Peano-continuum that contains no simple closed curve.
Here, a Peano-continuum is a continuum with locally-connected.
The crucial difference between a topological tree and a topological dendrite is that a topological dendrite is no necessary to become a finite graph.
In other words, a dendrite can contain the structure composed of infinitely many arcs and the self-similar structure.
For instance, the fern structure that has a self-similarity can be characterized by a topological dendrite. 
Figure \ref{Fig.SSdend} shows an example of a topological dendrite.
In this figure, the topological dendrite is composed of the infinitely many arcs of $F_i$ with the end points $p_i$ and $l_i$ $(i=1,2,\dots)$ and their convergence point $p_0$. 
Note that since any graph is a Peano-continuum, any tree becomes a dendrite.
In the practical branching (dendritic) pattern found in nature,  we can make use of either a topological tree or a topological dendrite for suiting the pattern.

\subsection{Direct Sum (Clusterized Sturcture)}
\label{sec:2-3}

\begin{figure}[h!]
    \begin{center}
	\includegraphics[width=4cm]{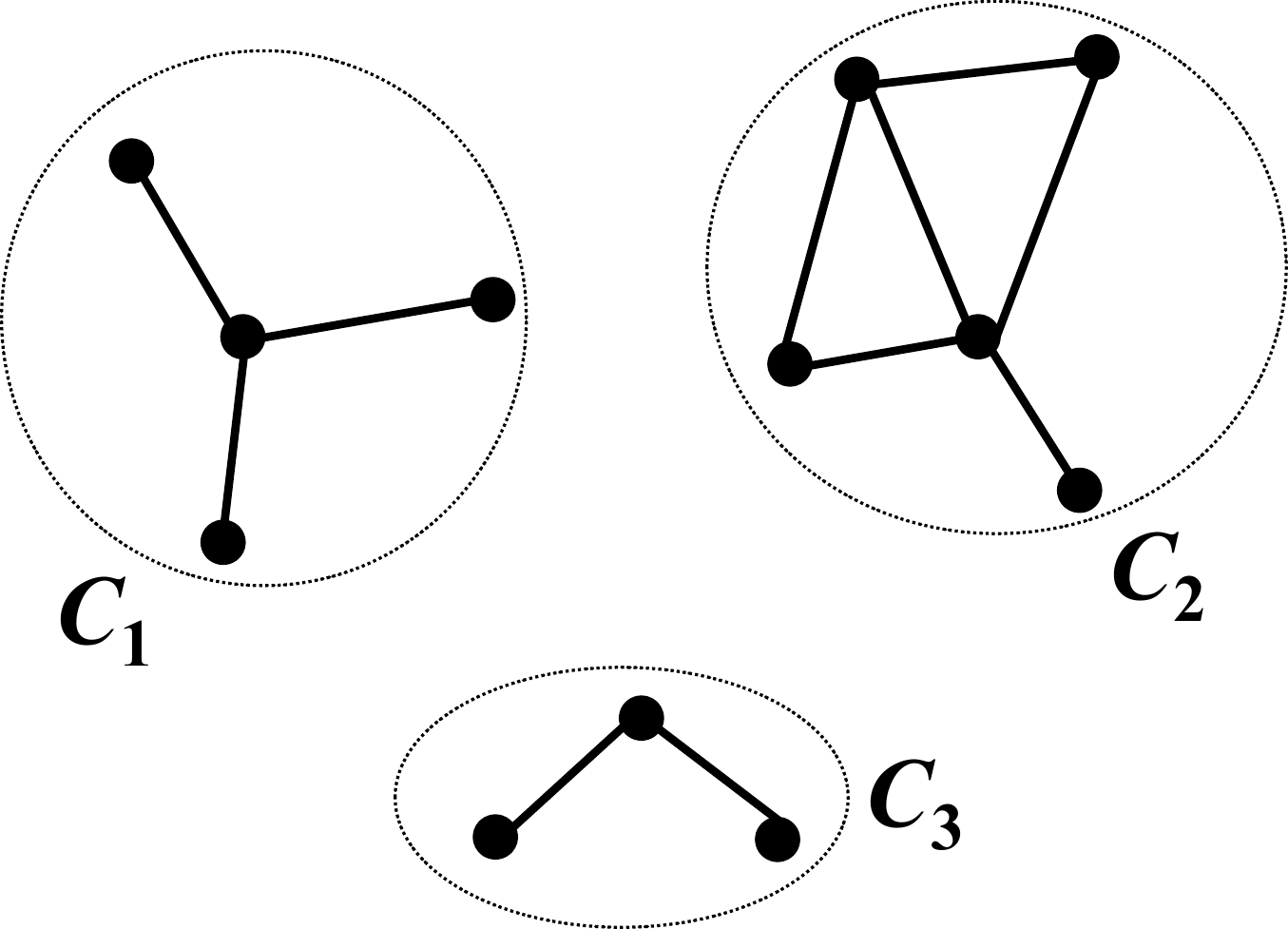}
    \hspace{13mm}
    \includegraphics[width=3.5cm]{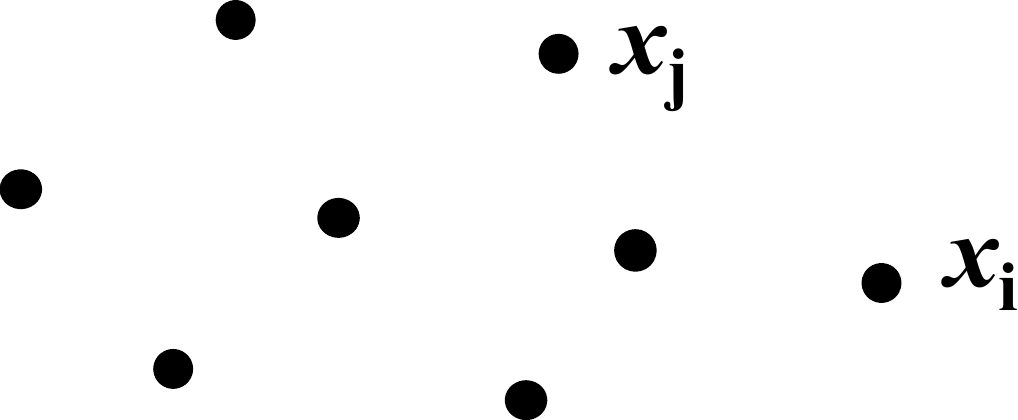}\\
    (a)\hspace{5cm}(b)
    \caption{\label{Fig.cluster} 
    Direct sums of
    (a) the three graphs $C_1,C_2$ and $C_3$, and 
    (b) the family of one-point continuum $\{x_i\}$.
   }
   \end{center}
\end{figure}

For describing the pattern with a clusterized structure,
a topological space called the direct sum is suitable.
For the given family of topological spaces $\{(X_i,\tau_{i});i=1,\dots,s\}$ satisfying $X_i\cap X_j = \emptyset, i\not = 
j$, the direct sum $\Big{(}\bigoplus_{i=1}^sX_i,\bigoplus_{i=1}^s\tau_i\Big{)}$ of them is defined as the space $
\bigoplus_{i=1}^sX_i=\bigcup _{i=1}^sX_i$ that equips the topology $\bigoplus_{i=1}^s\tau_i=\Big{\{}U\subset \bigoplus_{i=1}^sX_i;U\cap X_i\in 
\tau_{i},i=1,\dots,s\Big{\}}$\cite{Engelking}. 
The direct sum is always disconnected.
By corresponding each cluster to $X_i$, the direct sum provides the whole description of the set of the clusters as one topological space.
For instance, the clusterized structure whose cluster possesses a network pattern is characterized by the the direct sum $\bigoplus_{i=1}^sX_i$ of the finite graphs $X_i$.  
Figure \ref{Fig.cluster} (a) shows this case.
By replacing a graphic structure with other continua such as a tree and a dendrite, their clusterized structure can be also constructed.
Note that for the continua composed of just one point continuum $\{x_i\}$ $(i=1,\dots,s)$ its direct sum becomes $\{x_1,\dots,x_s\}$, which is a finite totally disconnected space (Fig. \ref{Fig.cluster} (b)).
Furthermore, we can consider a direct sum $X=\bigoplus_{i=1}^n X_i$ of the clusters $X_i$ $(i=1,\dots, n)$ where 
each $X_i$ is also the direct sum of the other clusters, $X_i=\bigoplus_{k=1}^{s(i)} X_{i,k}$.
Then, the direct sum $X$ characterizes a hierarchic structure of the clusters.

%
%

It is noted that the property, the topological spaces appearing in the above classifications are every compact metric spaces, is crucial.
Actually, in order to gain the topological representations of any pattern, we utilize the mathematical fact that any compact metric space is homeomorphic to a decomposition space of a 0-dim, perfect, compact Hausdorff-space. 
In the next section, a practical construction of such topological representation is shown based on a Cantor set.
%

\section{Topological Self-similarity of Cantor set}
\label{sec:3}

In this section, we focus on the Cantor set and demonstrate that it has a topological self-similarity, which is characterized by a pair of contractions.
The obtained self-similar property is used to construct a topological representation. 

At first, we recall what topological self-similar is\cite {Hata1985}.
A topological space $(A,\tau)$ is topologically self-similar if it satisfies the following conditions; 
(i) $(A,\tau)$ is metrizable, namely, the topology $\tau$ is identical with a metric topology $\tau_d$ specified by a metric $d$ on $A$ and (ii) a set of contractions $\{f_j:(A,\tau_d)\to (A,\tau_d), j=1,\dots, m (2\leq m <\infty)\}$ that satisfies 
\begin{eqnarray}
    \cup_{j=1}^mf_j (A)=A
    \label{eqn:2-0}
\end{eqnarray} 
is found.
Here, a map $f:(A,\tau_d)\to (A,\tau_d)$ on a metric space $(A,\tau_d)$ is a contraction if there is a constant $0\leq \alpha<1$ such that $d(f(x),f(y))\leq \alpha d(x,y)$ $(x,y\in A)$.
The relation (\ref{eqn:2-0}) shows the self-similar fractal property of $A$ and hence the topological self-similarity is considered as a mathematical generalization of self-similar fractal sets. 
Indeed, several familiar self-similar sets such as CMTS, Sierpinski gasket, Koch curve, and Menger sponge are topological self-similar.

Next we set a Cantor set.
Let $\{0,1\}^\mathbb{N}$ be the set of infinitely countable product of $\{0,1\}$ whose topology
$\tau_d $ is a metric topology with the metric defined by $d(x, x')=\sum_{i=1}^{\infty}|x _i-x' _i|/2^i ~(x,x' \in \{0,1\}^{\mathbb{N}})$.
Note that $(\{0,1\}^\mathbb{N},\tau_d )$ is a 0-dim, perfect, compact metric space.
%
%
And $(\{0,1\}^\mathbb{N},\tau_d )$ is homeomorphic to CMTS$=C$.
Indeed, set a function $h$ from $\{0,1\}^\mathbb{N}$ onto $C$ by
	\begin{eqnarray}
		h : \{0,1\}^\mathbb{N} \to C, x \mapsto \sum_{i=1}^\infty \frac{2}{3^i}x_i~~~(x=(x_1,x_2,\dots) \in \{0,1\}^\mathbb{N}).
		\label{eqn:2-1}
	\end{eqnarray}
Then, it becomes a homeomorphism.
Thus, $(\{0,1\}^\mathbb{N},\tau_d )$ is a Cantor set.
To show its self-similarity, we set
	\begin{eqnarray}
		\begin{cases}
    	F_0 : \{0,1\}^\mathbb{N} \to \{0,1\}^\mathbb{N}, x=(x_1,x_2,\dots) \mapsto F_0(x)=(0,x_1,x_2,\dots),\\
    	F_1 : \{0,1\}^\mathbb{N} \to \{0,1\}^\mathbb{N}, x=(x_1,x_2,\dots) \mapsto F_1(x)=(1,x_1,x_2,\dots).
		\end{cases}
	\label{eqn:2-2}
	\end{eqnarray}
It is confrimed that $F_0$ and $F_1$ are injective and the following relations hold : for each $i=0,1,$
    \begin{eqnarray}
	d(F_i(x),F_i(x')) & = & \frac{1}{2}d(x, x')~~~(x,x' \in \{0,1\}^\mathbb{N}),
	\label{eqn:2-3}
	\end{eqnarray}
and
	\begin{eqnarray}
	F_0(\{0,1\}^\mathbb{N})\cup F_1(\{0,1\}^\mathbb{N}) = \{0,1\}^\mathbb{N}.
	\label{eqn:2-4}
	\end{eqnarray}
%
%
%
%
The relation (\ref{eqn:2-3}) shows that $F_0$ and $F_1$ are contractions on $\{0,1\}^\mathbb{N}$ in terms of the metric $d$.
In addition, from (\ref{eqn:2-4}), it is found that $\{0,1\}^\mathbb{N}$ has a topological self-similar structure.
%
%

For characterizing a geometric pattern $Y$ based on the Cantor set $\{0,1\}^\mathbb{N}$, it is necessary to find a continuous map from $\{0,1\}^\mathbb{N}$ onto $Y$.
The continuous map can be obtained, methodically, by decomposing $\{0,1\}^\mathbb{N}$ into its cones, where a subset forming $\{k_1\}_{1}\times \cdots \times \{k_i\}_{_{i}}\times \{0,1\}^{\mathbb{N} -\{1,\dots, i\}} = \{x : \mathbb{N} \to \{0,1\}, x(l)=k_l\in \{ 0,1\}, l=1,\dots,{i}\}$ of $\{0,1\}^\mathbb{N}$ is called a cone. 
Note that the decomposition can be performed generally when the space is 0-dim perfect T$_0$-space.
The cone is expressed by a finitely composition of the contractions $F_0$ and $F_1$, 
	\begin{eqnarray}
	\{k_1\}_{1}\times \cdots \times \{k_i\}_{_{i}}\times \{0,1\}^{\mathbb{N} -\{1,\dots, i\}} = F_{k_1}\circ \dots \circ F_{k_i}(\{0,1\}^\mathbb{N}),
	\label{eqn:2-5}
	\end{eqnarray}
where $k_l \in \{0,1\}~(l=1,\dots, i <\infty)$.
This representation (\ref{eqn:2-5}) of the cone using the contractions $F_0$ and $F_1$ is called the ``contraction-cone'' of $\{0,1\}^{\mathbb{N}}$, hereafter.
Note that each contraction-cone is also a Cantor set.
By using the contraction-cones (\ref{eqn:2-5}), $\{0,1\}^{\mathbb{N}}$ is decomposed as $\{0,1\}^{\mathbb{N}}=\oplus_{i=1}^n X_i$, where
\begin{equation}
	\left\{
		\begin{array}{lcl}
		X_1=F_0(\{0,1\}^{\mathbb{N}}),\\
		X_i=F_1\circ \cdots \circ F_1 \circ F_0(X)=F_1^{i-1}\circ F_0(\{0,1\}^{\mathbb{N}}) ~(i=2,3,\dots,n-1),\\
		X_{n}=F_1^{n-1}(\{0,1\}^{\mathbb{N}}).
		\end{array}
	\right.
\label{eqn:2-7}
\end{equation}
%
%
%
%
Each $X_i$ is also decomposed by a set  $\{X_{i_1},\dots,X_{i_{n_i}}\}$ of the contraction-cones where 
\begin{equation}
	\left\{
		\begin{array}{lcl}
		X_{i_1} = F_1^{i-1}\circ F_0 \circ F_0(\{0,1\}^{\mathbb{N}}) = F_1^{i-1}\circ F_0^2(\{0,1\}^{\mathbb{N}}),\\
		X_{i_j} = F_1^{i-1}\circ F_0 \circ F_1^{j-1} \circ F_0^j (\{0,1\}^{\mathbb{N}}) ~(j=2,3,\dots,n_i-1),\\
		X_{i_{n_i}} = F_1^{i-1}\circ F_0 \circ F_1^{n_i-1} (\{0,1\}^{\mathbb{N}}).
\end{array}
	\right.
\label{eqn:2-8}
\end{equation}
Continuing the decomposing procedure provides a continuous map $f : \{0,1\}^{\mathbb{N}} \to Y$\cite{Ohmori2019}.
By using the continuous onto map $f$, the decomposition space $\mathcal{D}_f=\{ f^{-1}(y);y\in Y \}$ with a decomposition topology $\tau(\mathcal{D}_f)=\{\mathcal{U}\subset \mathcal{D}_f; \bigcup \mathcal{U} \in \tau_d\}$ of $\{0,1\}^{\mathbb{N}}$ is obtained to be homeomorphic onto $Y$. 
This decomposition space $\mathcal{D}_f$ becomes the topological representation of the geometric pattern $Y$.

As the simplest case, we see the  topological representation of a closed interval $[0,1]$.
Note that this representation is immediately available to obtain the representation of the arc pattern because an arc is homeomorphic to $[0,1]$.
First the following relation that is a generalization of the relation (\ref{eqn:2-5}) is satisfied : for $k_1,k_2,\dots \in\{0,1\},$
	\begin{eqnarray}
	\{k_1\}_{1}\times \{k_2\}_{2} \times \cdots \times \{0,1\}^{\mathbb{N} -\{1,2,\cdots \}} & = & \lim F_{k_1}\circ F_{k_2}\circ \dots \circ F_{k_n}(\{0,1\}^{\mathbb{N}}) \nonumber \\
	& = & \{(k_1,k_2,\dots)\} ~~\text{(singleton)},
	\label{eqn:2-9}
	\end{eqnarray}
where $\lim$ stands for the limit on $\Im(\{0,1\}^{\mathbb{N}})-\{ \emptyset\}$, $\Im(\{0,1\}^{\mathbb{N}})$ is the family of closed sets of $\{0,1\}^{\mathbb{N}}$, and the first equation of the right hand side expresses the limit for the sequence $\{F_{k_1}(\{0,1\}^{\mathbb{N}}), F_{k_1}\circ F_{k_2}(\{0,1\}^{\mathbb{N}}), \dots \}$ in $\Im(\{0,1\}^{\mathbb{N}})-\{ \emptyset\}$ 
\footnote{
    For a sequence $\{A_n\}_{n\in \mathbb{N}}$ of $2^{\{0,1\}^\mathbb{N}}$, $\lim A_n \equiv \lim \inf A_n = \lim \sup A_n$ 
    where $\lim \inf A_n = \{x\in \{0,1\}^\mathbb{N} ;$ for each open set $U$ of $\{0,1\}^\mathbb{N}$ containing $x$, $U\cap A_n\not = \emptyset$ for all but finitely many $n \}$ and $\lim \sup A_n =\{x \in \{0,1\}^\mathbb{N} ;$ for each open set $U$ of $\{0,1\}^\mathbb{N}$ containing $x$, $U\cap A_n \not = \emptyset$ for infinitely many $n \}$. 
%
%
Note that $\lim (A_n \cup B_n)= \lim A_n \cup \lim B_n$. 
    The relation (\ref{eqn:2-9}) is obtained as follows; it is clear that $\{k_1\}_{1}\times \{k_2\}_{2} \times \cdots \times \{0,1\}^{\mathbb{N} -\{1,2,\cdots \}} \subset \lim \inf F_{k_1}\circ F_{k_2}\circ \dots \circ F_{k_n}(\{0,1\}^{\mathbb{N}})$. 
    If $x \not \in \{k_1\}_{1}\times \{k_2\}_{2} \times \cdots \times \{0,1\}^{\mathbb{N} -\{1,2,\cdots \}}, x_{i_0}\not = k_{i_0}$ for some $i_0\in \mathbb{N}$. 
    Letting an open set $u = \{x_{i_0}\}_{i_0}\times \{0,1\}^{\mathbb{N}-\{i_0\}}$ of $\{0,1\}^\mathbb{N}$ containing $x$, it follows that for any infinite countable index set $I$, there exists $l\in I$ with $l>i_0$ such that $u \cap F_{k_1}\circ \cdots \circ F_{k_{l}}(X) = \emptyset$. 
    Therefore, $\lim \sup F_{k_1}\circ F_{k_2}\circ \dots \circ F_{k_n}(\{0,1\}^{\mathbb{N}}) \subset \{k_1\}_{1}\times \{k_2\}_{2} \times \cdots \times \{0,1\}^{\mathbb{N} -\{1,2,\cdots \}}$. 
For the detailed discussion about the limits, see A. Illanes and S. B. Nadler Jr., \emph{Hyperspaces} (Marcel Dekker, New York, 1999). 
}.
For the Cantor cube model $\{0,1\}^{\Lambda}$, the topological representation of $[0,1]$ is obtained as follows;
(i) when $y=\Sigma _{i=1}^\infty a_i/2^i \not \in M$ for some $a_1,a_2,\dots\in \{0,1\}$, where $M\equiv \{l/2^n ; n=1,2,\dots$ and $ l=1,\dots,2^n-1\}$,
\begin{eqnarray}
		f^{-1}(y)=\{a_1\}_{\lambda _1}\times \{a_2\}_{\lambda _2}\times \cdots \times \{0,1\}^{\Lambda -\{\lambda_1,\lambda_2,\cdots\}}, 
		\label{eqn:2-10-01}
\end{eqnarray}
(ii) when $y = l/2^n\in M$,
\begin{eqnarray}
		f^{-1}(y)=\Big{[} \{a_1\}_{\lambda _1}\times \{a_2\}_{\lambda _2}\times \cdots \times \{a_{n-1}\}_{\lambda _{n-1}}\times \{0\}_{\lambda _{n}}\times \{1\}_{\lambda _{n+1}}\times \{1\}_{\lambda _{n+2}}\times  \cdots \times \{0,1\}^{\Lambda -\{\lambda _1,\lambda_2,\cdots\}}\Big{]}
		\nonumber \\
		\cup \Big{[}\{a_1\}_{\lambda _1}\times \{a_2\}_{\lambda _2}\times \cdots \times \{a_{n-1}\}_{\lambda _{n-1}}\times \{1\}_{\lambda _{n}}\times \{0\}_{\lambda _{n+1}}\times \{0\}_{\lambda _{n+2}}\times  \cdots \times \{0,1\}^{\Lambda -\{\lambda _1,\lambda_2,\cdots\}} \Big{]}	
		\label{eqn:2-10-02}
\end{eqnarray}
for some $a_1,\dots, a_{n-1}$, where $\lambda_1,\lambda_2,\dots \in \Lambda $. 
Applying (\ref{eqn:2-9}) to (\ref{eqn:2-10-01}) and (\ref{eqn:2-10-02}), the topological representation  of $[0,1]$ is reconstructed as follows; (i) for  $y=\Sigma _{i=1}^\infty a_i/2^i \not \in M$
\begin{eqnarray}
		f^{-1}(y)=\lim F_{a_1}\circ F_{a_2}\circ \dots \circ F_{a_n}(\{0,1\}^{\mathbb{N}})=\{(a_1,a_2,\dots)\}, 
		\label{eqn:2-10-1}
\end{eqnarray}
and (ii) for $y = l/2^n\in M$
\begin{eqnarray}
		f^{-1}(y)=\lim F_{a_1}\circ F_{a_2}\circ \dots \circ F_{a_{n-1}}\circ (F_1\circ F_0^{n}(\{0,1\}^{\mathbb{N}}) \oplus F_0\circ F_1^{n}(\{0,1\}^{\mathbb{N}}))=\{a^0,a^1\}
		\label{eqn:2-10-2}
\end{eqnarray}
for some $a_1,\dots, a_{n-1}$, where $a^0=(a_1,\dots,a_{n-1}, 0, 1, 1,\dots)$ and $a^1=(a_1,\dots,a_{n-1}, 1, 0, 0,\dots)$. 
Here, $f^{-1}(0)=\{e_0\}$ and $f^{-1}(1)=\{e_1\}$ where $e_0 = (0,0,\dots)$ and $e_1=(1,1,\dots)$ are the elements of $\{0,1\}^\mathbb{N}$.

\section{Topological representations of geometric patterns}
\label{sec.4}

The method of topological representation obtained in the previous section is now applied to the topological spaces shown in Sec. 2.
%


%
First of all, we consider an arc $E$  with two end points $e_1$ and $e_2$ as shown in (b) of Fig. \ref{Fig.graphs}.
Setting a homeomorphism $h$ from $E$ to $[0,1]$,
a point $a \in E$ is mapped into $h(a)\in [0,1]$.
With the aid of the representations  (\ref{eqn:2-10-1}) and (\ref{eqn:2-10-2}) of $[0,1]$, we can obtain the topological representation of $E$ based on the contractions $F_0$ and $F_1$ of the Cantor set $\{0,1\}^{\mathbb{N}}$, as follows; 
if $h(a) \not \in M (\equiv \{l/2^n ; l=1,\cdots,2^n-1, n=1,2,\cdots\})$, $h(a)$ satisfies $h(a) = \Sigma _{i=1}^{\infty} k_i/2^i$
 for some $k_1, k_2, \dots \in \{0,1\}$ and then
\begin{eqnarray}
      a \doteq  \lim F_{k_1}\circ F_{k_2}\circ \dots \circ F_{k_n}(\{0,1\}^{\mathbb{N}})=\{(k_1,k_2,\dots)\}.
\label{eqn:3-1-1}
\end{eqnarray}
If $h(a) \in M$, then
\begin{eqnarray}
      a  \doteq \lim F_{k_1}\circ F_{k_2}\circ \dots \circ F_{k_{m-1}}\circ (F_1\circ F_0^{m}(\{0,1\}^{\mathbb{N}}) \oplus F_0\circ F_1^{m}(\{0,1\}^{\mathbb{N}}))=\{k^0,k^1\},
\label{eqn:3-1-2}
\end{eqnarray}
for some $m$ and some $k_1,\dots, k_{m-1} \in \{0,1\}$ where $k^0=(k_1,\dots,k_{m-1}, 0, 1, 1,\dots)$ and $k^1=(k_1,\dots,k_{m-1}, 1, 0, 0,\dots)$.
Here, $\doteq$ stands for the sign of identification of $a$ with a corresponding point $f^{-1}(a)$ of the decomposition space. 
%
%
By introducing the sign $L_a$ defined as, 
\begin{equation}
L_a \equiv 
	\begin{cases}
	(\ref{eqn:3-1-1}), & h(a) \not\in M,\\
	(\ref{eqn:3-1-2}), & h(a) \in M,
	\end{cases}
\label{eqn:3-2}
\end{equation}
the representations (\ref{eqn:3-1-1}) and (\ref{eqn:3-1-2}) can be simplified as 
\begin{eqnarray}
      a \doteq L_a ~~~(a\in E).
\label{eqn:3-3}
\end{eqnarray}
%
%
%
Note that assuming $h(e_1)=0$ and $h(e_2)=1$, the end points $e_1$ and $e_2$ form $e_1 \doteq \lim F_0^n(\{0,1\}^{\mathbb{N}})$ and $e_2 \doteq \lim F_1^n(\{0,1\}^{\mathbb{N}})$, respectively.
The representations (\ref{eqn:3-1-1}) and (\ref{eqn:3-1-2}) provide that each geometric element $a$ of the arc $E$ is described as the limit point of the composition $F_0$ and $F_1$.  
%

%
Next, we focus on a finite graph.
Set $Y_g$ to be a finite graph composed of $r$ numbers of arcs $E_1,\dots, E_r~~(r<\infty)$.
To associate with each of arcs $E_i$,
we construct a partition $\{X^1, \dots, X^r\}$ of $\{0,1\}^{\mathbb{N}}$ composed of the following contraction-cones :
\begin{equation}
	\left\{
		\begin{array}{lcl}
		X^1 = F_0(\{0,1\}^{\mathbb{N}}),\\
		X^i=F_1\circ \cdots \circ F_1 \circ F_0(\{0,1\}^{\mathbb{N}})=F_1^{i-1}\circ F_0(\{0,1\}^{\mathbb{N}}) ~(i=2,3,\dots,r-1),\\
		X^{r}=F_1^{r-1}(\{0,1\}^{\mathbb{N}}).
		\end{array}
	\right.
\label{eqn:3-4}
\end{equation}
%
%
%
%
For each arc $E_i$, the topological representation is obtained as the form (\ref{eqn:3-3}).
Considering this fact, the topological representation of $Y_g$ is 
\begin{eqnarray}
      y \doteq 
		F_1 ^{i-1} \circ F_0 (L^i_y)	
		\label{eqn:3-5}
\end{eqnarray}
for a point $y$ in an arc $E_i$,
where $L^i_{y}$ is defined by (\ref{eqn:3-2}) for the homeomorphism $h_i$ from $E_i$ onto $[0,1]$ instead of $h$, and 
\begin{eqnarray}
      x \doteq \cup _{j=1}^q F_1^{t_j-1} \circ F_0(L^{t_j}_{x})
		\label{eqn:3-6}
\end{eqnarray}
%
%
%
for a node $x$ connecting with arcs $E_{t_1},\cdots , E_{t_q}$.
In (\ref{eqn:3-5}) and (\ref{eqn:3-6}), $L^{i}_{y}$ and $L^{t_j}_{x}$ are the terms that represent the position of $y$ and $x$ in the arcs $E_i$ and $E_{t_j}$, respectively. 
$F_1^{i-1}\circ F_0$ shows that $y$ is contained only in $E_i$, whereas,  $\cup _{j=1}^q F_1^{t_j-1} \circ F_0$ shows that $x$ is the node connecting with the arcs $E_{t_1},\dots , E_{t_q}$. 
%
%
%
%
Note that (\ref{eqn:3-5}) and (\ref{eqn:3-6}) reproduce the representations of  (\ref{eqn:1-1}) in Sec. 1, which is obtained by the Cantor cube model $\{0,1\}^\Lambda$.
Indeed, it is found that for $z\in Y_g$, the factors $S^l_z$ and $X^l$ $(l=1,\dots,r)$ in (\ref{eqn:1-1}) correspond to $L^l_z$ and $F_1^{l} \circ F_0$ and therefore $S^l_z$ and $X^l$ can be characterized by the composition of $F_0$ and $F_1$.
To obtain the topological representation for the Cantor set $\{0,1\}^\mathbb{N}$ from Cantor cube model $\{0,1\}^\Lambda$,
therefore,
the transformations 
\begin{equation}
S^l_z \to L^l_z,~~~X^l \to F_1^{l} \circ F_0
\label{eqn:3-tra}
\end{equation}
are applicable. 
Then, the topological representations of a tree or a topological dendrite for $\{0,1\}^\mathbb{N}$ are also obtained from them for $\{0,1\}^\Lambda$ shown in the previous study\cite{Ohmori2019}.

The clusterized structure of network configuration can be realized by the topological space of the direct sum of finite graphs.
We focus on this structure and denote it by  $Y_c=(\bigoplus_{i=1}^sC_i, \bigoplus_{i=1}^s \tau _i)$, where $(C_i,\tau_i)$ is a finite graph composed of 
$E^{i}_1, \dots, E^{i}_{r(i)}$~$(r(i)<\infty)$.
%
%
To the disjoint clusters $C_1,\dots, C_s$, the contraction-cones $J_{1},\dots, J_{s}$ can be assigned, where
\begin{equation}
	\left\{
		\begin{array}{lcl}
		J_1 = F_0(\{0,1\}^{\mathbb{N}}),\\
		J_i=F_1^{i-1}\circ F_0(\{0,1\}^{\mathbb{N}}) ~(i=2,3,\dots,s-1),\\
		J_{s}=F_1^{r-1}(\{0,1\}^{\mathbb{N}}).
		\end{array}
	\right.
\label{eqn:3-7}
\end{equation}
%
%
%
Each finite graph $C_i$ has the topological representation of the forms (\ref{eqn:3-5}) and (\ref{eqn:3-6}).
Therefore, we obtain the representation for the whole space $Y_c$ as follows;
assuming that $a\in Y_c$ belongs to a cluster $C_{i_0}$, then
\begin{equation}
		a \doteq F_1^{i_0-1}\circ F_0 \circ
		\left\{
		\begin{array}{lcl}
		\cup _{j=1}^q F_1^{t_j-1} \circ F_0 (L^{t_j}_{a}), \\
		F_1 ^{i-1} \circ F_0(L^i_a),
		\end{array}
	\right.
\label{eqn:3-8}
\end{equation}
where $a$ is located either in the node connecting with the arcs $E_{t_1}^{i_0},\dots,E_{t_q}^{i_0}$~$(t_q\leq r(i_0))$ or in the arc $E^{i_0}_i$.
In the right hand side of (\ref{eqn:3-8}), the first composition of the contractions $F_1^{i_0-1}\circ F_0$ is the term that indicates a graph $C_{i_0}$ to which $a$ belongs, and the successive compositions gives the positioning of $a$ in the graph $C_{i_0}$.

For the special case where each cluster is composed of just one point, $C_i=\{ x_i \}$, shown in Fig. \ref{Fig.cluster} (b), we have $\bigoplus_{i=1}^sC_i = \cup_{i=1}^s\{x_i\}=\{x_1,\dots,x_s\}$.
Then, by using the contraction-cones $J_i$ ($j=1,\cdots,s$) of (\ref{eqn:3-7}), the topological representation is obtained as
%
%
\begin{eqnarray}
		x_i \doteq J_i ~(i=1,\dots, s).
		\label{eqn:3-9}
\end{eqnarray}

\begin{figure}[h!]
    \begin{center}
	\includegraphics[width=5cm]{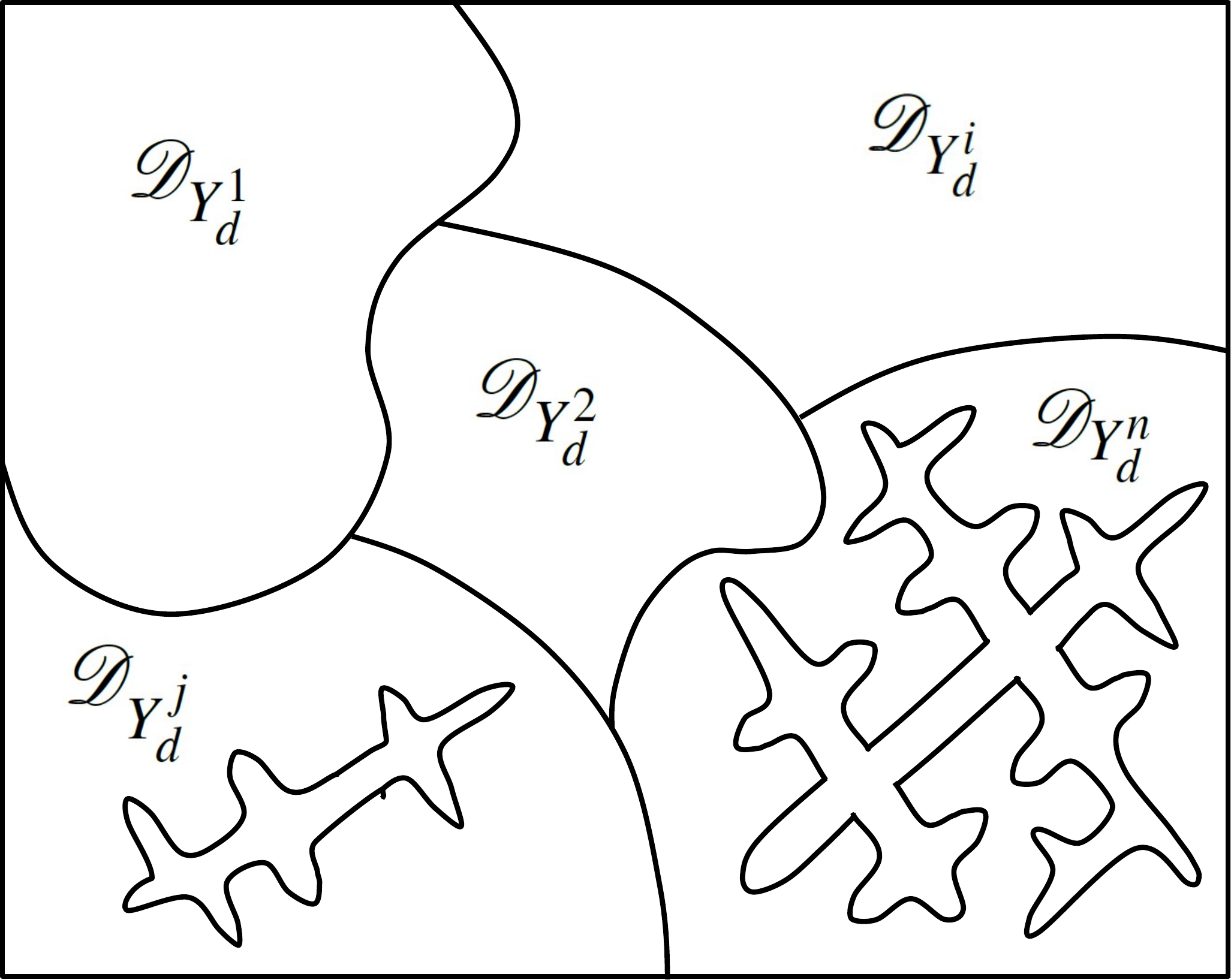}
    \hspace{13mm}
    \includegraphics[width=4cm]{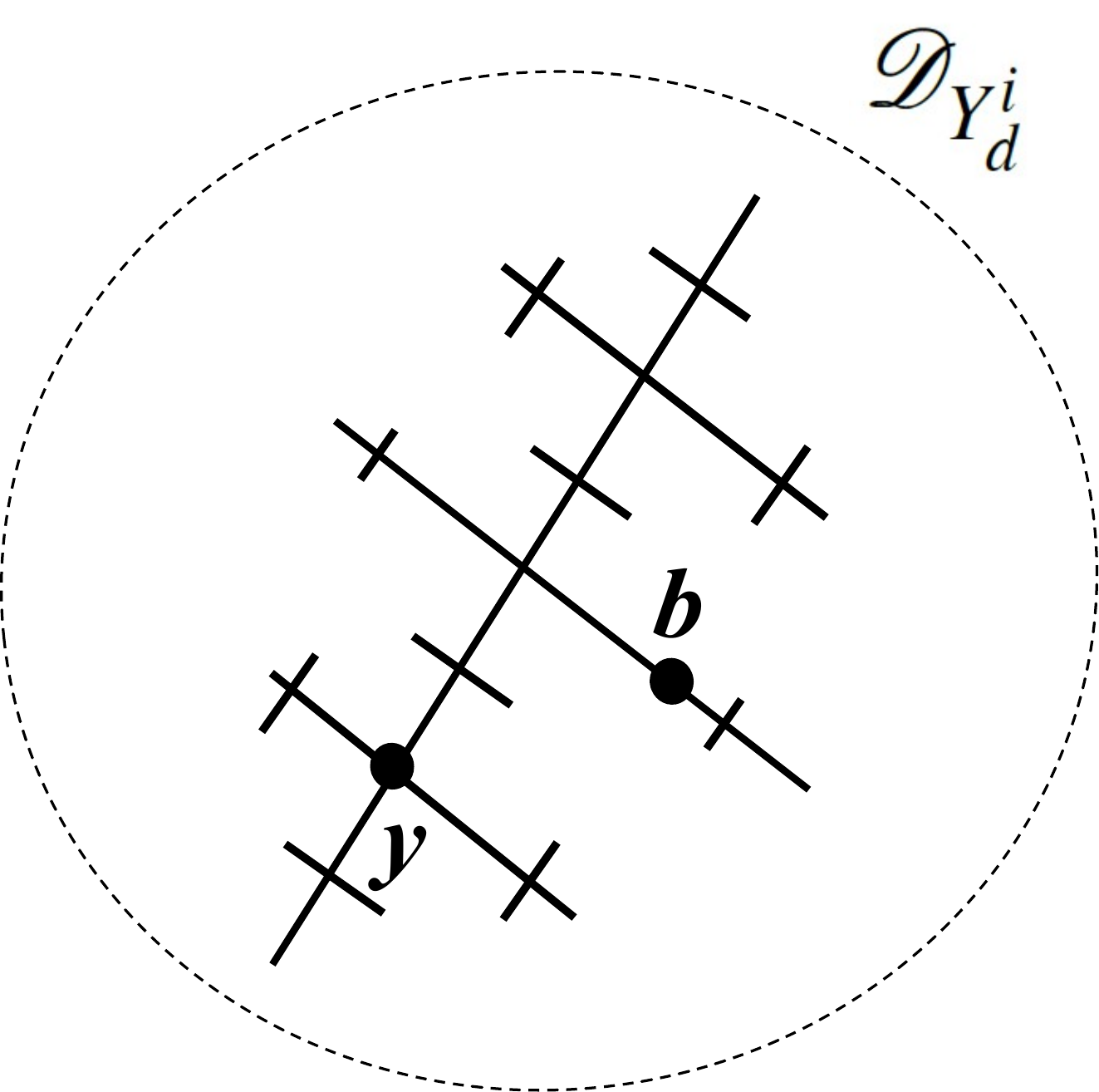}\\
    (a)\hspace{6cm}(b)
    \caption{\label{Fig.polycrystal} 
    Schematic explanations of (a) a polycrystal tiled by dendritic single crystals and (b) each dendritic single crystal that is identified with a tree cluster.
    $\mathcal{D}_{Y_d^i}$ is given as the relation (\ref{eqn:3-pory3}) that represents a tree cluster where $y$ is a node and $b$ is a point in an arc. 
   }
   \end{center}
\end{figure}

As an application of the topological representation for the clusterized pattern, 
we now consider a polycrystal filled with an $n$ number of single crystals whose single crystal is supposed to form a specific geometric pattern, e,g., dendritic pattern.
Such a situation is often found in solidification\cite{Porter}.
Fig. \ref{Fig.polycrystal} (a) shows the sketch of such polycrystal where the dendritic structure characterizes its each single crystal.
The topological representation of this type has been discussed in our studies, so far\cite{Kitada2016,Ohmori2019}.
In particular, by identifying this polycrystal with a kind of clusterized structure whose cluster is a tree (Fig. \ref{Fig.polycrystal} (b)),
we derived the topological representation using Cantor cube model, as follows. Each single crystal $Y_d^i (i=1,\dots, n)$ is represented by  
%
	$  \mathcal{D}_{Y_d^i} = \Big{\{} y \doteq J_i \cap \cup _{j=1}^q (X^{t_j} \cap S^{t_j}_{y}) ; y\in Y_d^i \Big{\}} \cup \Big{\{} b \doteq J_i \cap X^{j} \cap S^j_{b} ; b\in Y_d^i  \Big{\}} 
$
%
and their polycrystal is represented as the union of $\mathcal{D}_{Y_d^i}$, 
\begin{eqnarray}
		\mathcal{D} = \cup_{i=1}^n \mathcal{D}_{Y_d^i}.
		\label{eqn:3-pory2}
\end{eqnarray}
Note that $\mathcal{D}_{Y_d^i}$ and $\mathcal{D}_{Y_d^j}$ are mutually disjoint for $i\not =j$.
With the aid of the transformations (\ref{eqn:3-tra}) and the contraction-cones (\ref{eqn:3-7}), $\mathcal{D}_{Y_d^i}$ is reproduced as
\begin{eqnarray}
	  \mathcal{D}_{Y_d^i} = \Big{\{} y \doteq F_1^{i-1}\circ F_0 \circ \big{(} \cup _{j=1}^q F_1^{t_j-1} \circ F_0 (L^{t_j}_{y}) \big{)}
    ; y\in Y_d^i \Big{\}} 
    \cup \Big{\{} b \doteq F_1^{i-1}\circ F_0 \circ F_1 ^{j-1} \circ F_0(L^j_b) 
    ; b\in Y_d^i  \Big{\}},
		\label{eqn:3-pory3}
\end{eqnarray}
for $i=1,\dots, n$.
(\ref{eqn:3-pory3}) provides the representation of the dendritic single crystal characterized by the contraction $F_0$ and $F_1$ of the Cantor set,
and (\ref{eqn:3-pory2}) is the relationship of the single crystals to a whole polycrystal composed of them.
%
%

%

%
As we have shown in this section, each geometric pattern can be represented topologically by the Cantor set and its contractions, 
independently of the detail properties of each matter.
The construction of these representations in our method is based on the topological and fractal properties of the Cantor set.
In fact, the Cantor set has the topological properties of a 0-dim, perfect, compact Hausdorff-space, which supply the partition composed of cones.
In addition, the contractions characterizing the self-similarity of the Cantor set make up a contraction-cone that is the key for representing each pattern topologically. 
So far, mathematical methods using topological and fractal viewpoints, such as the scaling method, have been contributed to the development of the study of pattern formation physics\cite{Feder,Cross}. 
Therefore, we believe that analyzing the Cantor set by our current method may have the potential to provide universal description of the geometric patterns of disordered matter.

\section{Conclusion}
\label{sec:5}

We have investigated mathematical description that expresses universally the geometric patterns found in disordered matters from the viewpoint of the general topology.
These geometric patterns can be characterized by the concept of continua, and 
focusing on the practical patterns such as a network, branching, and clusterized structures, 
we have successfully classified them under the topological graph, tree (or topological dendrite), and the direct sum of continua.
To such topological spaces, we have associated the topological representations obtained in the basis of a Cantor set $(\{0,1\}^{\mathbb{N}},\tau_d)$ with the contractions $F_0$ and $F_1$ characterizing the self-similarity. 
It is found that in the topological representations, each element of the geometric patterns can be identified with the limit of the compositions of $F_0$ and $F_1$. 
%
%
%
%

%
%
%


\bigskip

\noindent
{\bf acknowledgments}

The authors are grateful to Prof. S. Matsutani at Kanazawa University for useful comments and
encouragements.
This work was supported by Institute of Mathematics for Industry, Joint Usage/Research Center in Kyushu University. (FY2022 Workshop II “Geometry and Algebra in Material Science III” (2022a003)).


%

\end{document}